# Magnonic Combinatorial Memory

Mykhaylo Balynskyy and Alexander Khitun*

*Department of Electrical and Computer Engineering, University of California - Riverside, Riverside, California, USA 92521*

Correspondence to akhitun@engr.ucr.edu

**Abstract:** In this work, we consider a type of magnetic memory where information is encoded into the mutual arrangements of magnets. The device is an active ring circuit comprising magnetic and electronic parts connected in series. The electric part includes a broad-band amplifier, phase shifters, and attenuators. The magnetic part is a mesh of magnonic waveguides with magnets placed on the waveguide junctions. There are amplitude and phase conditions for auto-oscillations to occur in the active ring circuit. The frequency(s) of the auto-oscillation and spin wave propagation route(s) in the magnetic part depends on the mutual arrangement of magnets in the mesh. The propagation route is detected with a set of power sensors. The correlation between circuit parameters and spin wave route is the base of memory operation. The combination of input/output switches connecting electric and magnetic parts, and electric phase shifters constitute the memory address. The output of power sensors is the memory state. We present experimental data on the proof-of-the-concept experiments on the prototype with just three magnets placed on top of a single-crystal yttrium iron garnet $Y_3Fe_2(FeO_4)_3$ (YIG) film. The results demonstrate a robust operation with On/Off ratio for route detection exceeding 35 dB at room temperature. The number of propagation routes scales factorial with the size of the magnetic part. Coding information in propagation routes makes it possible to drastically increase the data storage density compared to conventional memory devices. MCM with just 25 magnets can store as much as $25! = 10^{25}$ bits. Physical limits and constraints are also discussed.

I.      Introduction

There is an urgent need in the increase of data-storage density of information storage devices in the current era of big data. The global data will grow to 175 zettabytes (ZB) by 2025 according to the International Data Corporation [1]. Conventional storage systems may become unsustainable due to their limited data capacity, infrastructure cost, and power consumption [2]. In the traditional process of improving the data-storage density (e.g., the number of bits stored per area), better performance is achieved by the miniaturization of the data-storage elements. It stimulates a quest for nanometer-size memory elements such as DNA-based [3] or sequence-defined macromolecules [4]. At the same time, memory architecture and the principles of data storage remain mainly unchanged.

As an example, we would like to refer to Random Access Memory (RAM). In Fig.1(A), there is shown a high-level picture of RAM organization. The core of the structure is an array of memory cells where each cell stores one bit of information. There is a variety of memory cells exploiting different physical phenomena/devices/circuits for data storage such as static RAM [5], dynamic RAM [6], and magnetic RAM [7]. In all cases, RAM contains multiplexing and demultiplexing circuitry for cell addressing. Typically, the addressing of a required cell is accomplished by transistors electrically connecting the cell on the selected row/column with the output circuit. The correlation between the cell address and cell state is the essence of data storage in RAM. It is illustrated in Fig.1(B). It is shown in a table, where the first column depicts the memory cell binary address. The second column in the table depicts the cell state (i.e., 0 or 1). *The maximum number of bits stored in conventional memory is limited by the number of memory cells.* This is the common property of all existing RAMs. Regardless of the physical mechanism of data storage (e.g., mechanical, electrical, magnetic), conventional memory devices make use of the individual states of the memory cells.

Here, we consider the possibility of building a fundamentally different data-storage device, where information is stored in the *signal propagation route*. It allows us to drastically increase the number of memory states compared to conventional memories. The rest of the work is organized as follows. In the next Section II, we describe the principle of operation of Magnonic Combinatorial Memory (MCM). Next, in Section III, we present the results of numerical modeling illustrating MCM operation. Experimental data obtained for the prototype are presented in section IV. The Discussion and Conclusions are given in sections V and VI, respectively.

II.     Material Structure and Principle of Operation

In order to explain the principle of operation of combinatorial devices, we start with an example of a two-path active ring circuit schematically shown in Fig.2(A). It consists of two parts referred to as the active part the and passive part. The active part includes a nonlinear broadband amplifier $G$, variable phase shifter $\Psi(V)$, and a controllable attenuator $A(V)$. It is assumed that these components are frequency-independent (i.e., provide the same amplification, the same phase shift, and the same attenuation for all frequencies). The magnitude of the phase shift and attenuation level can be adjusted by the applied voltage $V$. The broadband amplifier $G$ is the only source of energy in the system. The passive part consists of two paths, where each path has a bandpass frequency filter $L(f)$, phase shifter $\Delta(f)$, and a power sensor $P$. The filters and the phase shifters in the passive part are frequency-dependent (i.e., provide attenuation and phase shift depending on the frequency of the signal). There are two conditions for auto-oscillations to occur in an active ring circuit [8]:

$$G \times A \times L(f) \geq 1, \tag{1.1}$$

$$\Psi + \Delta(f) = 2\pi k, \ k = 1,2,3,\ldots \tag{1.2}$$

where $L(f)$ is the signal attenuation in the passive path(s), $k$ is an integer. The first equation (1.1) states the amplitude condition for auto-oscillations: the gain provided by the broadband amplifier should be sufficient to compensate for losses in the passive part and losses introduced by the attenuator. The second equation states the phase condition for auto-oscillations: the total phase shift for a signal circulating through the ring should be a multiple of $2\pi$. In this case, signals come in phase every propagation round. A more detailed explanation of the selective signal amplification in a multi-path active ring circuit can be found in Ref. [9].

There are four possible scenarios for the circuit shown in Fig.2(A). (i) There may be no auto-oscillations in the circuit as conditions (1.1) and/or (1.2) are not satisfied. (ii) There may be auto-oscillations in the circuit where the most of circulating power is coming through the top path. It happens when the two paths have different transmission/phase shift characteristics so the wave(s) on the resonant frequency can propagate only through the top waveguide but significantly attenuated in the bottom waveguide. (iii) The auto-oscillation may take place and most of the power is coming through the bottom waveguide. (iv) There may be auto-oscillations in the circuit with equal power flow through both paths in case of identical phase shifts and attenuation in the top and the bottom waveguides. As an example, let us consider a two-path active ring circuit as in Fig.2(A), where the bandpass filters on the top and the bottom paths are set to the different frequencies $f_1$ and $f_2$, respectively. The phase shifters provide the different phase shifts $\Delta_1 = 1.5\pi$ and $\Delta_2 = 0.5\pi$. It is assumed that the amplitude condition (1.1) is satisfied for all frequencies. There are two positions of the external phase shifter $\Psi$ at which the phase condition (1.2) is met. In the case $\Psi + \Delta_1 = 2\pi$, there are auto oscillations on frequency $f_1$ where the signal goes through the top waveguide. In the case $\Psi + \Delta_2 = 2\pi$, there are auto oscillations on frequency $f_2$ where the signal goes through the bottom waveguide. There are no auto oscillations for other positions of the external phase shifter $\Psi$. In Fig.2(B), the auto oscillation power is shown as a function of the external phase shifter $\Psi$. The insets illustrate the signal propagation route. The green circle corresponds to the power sensor detecting power. The grey circle depicts no power flow. The plot in Fig.2(B) corresponds to the circuit with ideal bandpass filters transmitting signals only on the central frequencies with zero bandwidth.

In our preceding experimental works, we demonstrated several electromagnetic active ring circuits comprising active electric and passive multi-path magnonic (spin wave) parts [9, 10]. There are several advantages of using spin waves in the active ring circuits. (i) Spin wave is a collective oscillation of a large number of spins in magnetic lattice, where a quantum of spin wave is called a magnon. The collective nature of spin wave reveals itself in a relatively long coherence length which may exceed millimeters at room temperature in ferrites [11]. Spin waves propagate much slower compared to electro-magnetic waves of the same frequency, which provides a significant phase shift even in the micrometer-length waveguides. (ii) The propagation of spin waves is affected by the local magnetic field produced by micro magnets placed on top of ferrite film. This property was used for magnetic bit read-out in a number of works [12-14]. *A micro magnet placed on top of ferrite film acts as a frequency filter and a phase shifter* *at the same time*. The bandpass frequency range and the phase shift can be engineered by the material, size, shape, and alignment of the magnet on top of ferrite waveguide.

The most appealing property of the active-ring circuit is the ability to find or self-adjust to the resonant frequency(s). The system starts with a superposition of signals of all possible frequencies. However, only signal on the resonant frequency(s) $f_r$ satisfying conditions (1.1) and (1.2) will be amplified. The propagation paths of the signals may differ depending on the arrangement of the frequency filters. This property of multi-path active ring circuits can be utilized for solving nondeterministic polynomial (NP) problems such as Prime Factorization [9], and traveling salesman problem [15]. In this work, we consider the multi-path active ring circuits for memory application.

The schematics of the proposed MCM are shown in Fig.3(a). It is an active ring circuit where the magnetic part consists of a mesh of magnonic waveguides. For simplicity, it is shown a 2D 5×5 mesh. The waveguides are shown in the light blue color. There are magnets depicted by the blue-red rhombs placed on top of waveguide junctions. It is assumed that each magnet can be in one of the eight thermally stable states corresponding to the eight different directions of magnetization with respect to the waveguide. The magnets can be of three different sizes. Thus, there are total 24 possible states for the junction with magnet plus one more state corresponding to the junction without a magnet. These magnets are aimed to act as the frequency filters and phase shifters for the propagating spin waves. Both the direction and the strength of the local magnetic field provided by the magnets are important for spin wave transport [16]. There are power sensors placed on top of the waveguides between the junctions. These sensors are aimed to detect the power of the spin wave signal and show the spin wave propagation route(s). For example, it may be Inverse Spin Hall Effect (ISHE) detectors. Being of a relatively simple material structure (e.g., Pt wire on top of YIG waveguide), ISHE provides output voltage proportional to the amplitude of spin wave [17]. There are $(2n-1) \times n$ detectors in the mesh with $n$ columns and $2n-1$ rows (e.g., 45 detectors shown in Fig.3(a)). Each sensor provides a voltage output $V_{ij}$, where subscripts $i$ and $j$ correspond to the row and column numbers, respectively.

The magnonic mesh is connected to the electric part via the input and output ports located on the left and right sides of the mesh. The conversion from electromagnetic wave to spin wave and vice versa may be accomplished with the help of micro antennas [18]. There is a switch (e.g., a transistor similar to one used in conventional memory for row/column addressing) at each input and output port to enable/disable the antenna for signal generation/receiving. These switches are aimed to control the combination of input/output antennas (e.g., input ports #2 and #3, output ports #1, #2, and #5). There are voltage-tunable frequency filters $f_i$, voltage-tunable phase shifters $\Psi_i$, and voltage-tunable attenuators $A_i$ at each output port, where the subscript $i$ depicts the output number. The phase shifters have discrete states corresponding to the specific phase shifts (e.g., $\Psi(0) = 0\pi$, $\Psi(1) = 0.25\pi$, $\Psi(2) = 0.5\pi$, etc.). The frequency filters and attenuators are analog devices to be used for system calibration/adjustment as will be described later in the text. There is one broadband amplifier $G$ in the electric part.

The principle of operation of MCM is based on the correlation between the combination of the input/output switches, phase shifters $\Psi_i$, and the spin wave propagation route in the mesh (e.g., the output voltages of the power detectors $V_{ij}$). The memory address includes a binary number corresponding to the states of input switches, a binary number corresponding to the states of output switches, and a binary number corresponding to the states of the phase shifters $\Psi_i$. The binary number for switches is an $n$-bit number, where 1 corresponds to the state On and 0 corresponds to the state Off.

For example, the mesh shown in Fig.3(a) has input ports #2, #3, and #4 in the position On. It corresponds to the binary address 01110. The output switches #1, #3, #4, and #5 are in the position On. It corresponds to the binary number 10111. There are more than two states for each phase shifter. In this case, the length of the binary number corresponding to the phase shifter states is related to $z^n$ possible combinations, where $z$ is the number of states per phase shifter. The memory state is the signal propagation route (i.e., the output voltages $V_{ij}$). One may introduce a reference voltage $V_{ref}$ to digitize the output. For instance, the output state is 1 if $V_{ij} \geq V_{ref}$, and 0 otherwise. In the example shown in Fig3.(a), the memory state is a sequence of 45 bits (one bit for each sensor).

The correlation between the memory addresses and the memory states is illustrated in Fig.3(b). It is shown an example of the truth table, where the first column is the memory address, and the second column is the memory state. The number of possible combinations of input switches is $2^n - 1$. It excludes one combination where all input ports are disconnected from the mesh. There is the same number of possible combinations of the output ports. In Fig.3(b), the first memory address corresponds to the case when the input port #1 and the output port #1 are connected to the mesh. The phase shifters are set to state 0. Every combination of switch states and phase shifters states constitutes an address. The total number of addresses is given as follows:

$$\# \, adresses = (2^n - 1)^2 \times z^n. \qquad (2)$$

There is one memory state that is related to the given address. The memory state is a sequence of 45 zeros and ones. In the left column of the truth table in Fig.3(b), these zeros and ones are arranged in nine rows with five columns with five bits per row. In this case, the position of bit reflects the position of the power sensor in the mesh. For example, the middle row in the truth table in Fig.3(b) corresponds to the case shown in Fig.3(a). The top five bits are zeros corresponding to the five power sensors located in the top row in the mesh. In general, there are $(2n - 1) \times n$ power sensors in the mesh and same number of bits per address. The total capacity of MCM is the product of the number of memory addresses and the number of bits stored per address that can be calculated as follows:

$$\# \, bits \, stored = (2^n - 1)^2 \times z^n \times n \times (2n - 1) \qquad (3)$$

According to Eq.(3), the data storage capacity of MCM scales according to the power law with the size of the mesh. It should be noted that the number of possible combinations of row/column switches for conventional RAM (e.g., as shown in Fig.1) also scales according to the power law. However, only a limited number of addresses store information (e.g., addresses including one row and one column). The states of other addresses (e.g., addresses with two rows and three columns) can be computed. The information in MCM is stored in the mutual arrangement of magnets in the mesh. There are $n!$ ways to have an ordered arrangement of $n$ distinct objects [19]. Considering a set of $n^2$ distinct magnets in the mesh with $n^2$ junctions, the number of ordered arrangements (permutations) is given by

$$\# \, ordered \, magnet \, arrangements \, = (n^2)! \qquad (4)$$

It is important that the number of possible magnet arrangements increases faster than the number of bits stored. It may be possible to find an arrangement of magnets so the spin wave propagation routes match the given truth table (e.g., shown in Fig.3(b)). For example, there are $25! = 1.55 \times 10^{25}$ possible arrangements for 25 magnets in $5 \times 5$ mesh as shown in Fig.3(a). That is the maximum number of bits that can be stored. The increase of the number of states per phase shifter above a certain value will increase the number of addresses but with identical states. Needless to say, that exploiting all the possible arrangements (routes) in $5 \times 5$ mesh will cover all needs in the current information storage (175 zettabytes $= 1.75 \times 10^{23}$ bits. In this part, we restricted our consideration by the read-out procedure to emphasize the fundamental enhancement in data storage density of MCM compared to traditional memories.

### III.     Results of numerical modeling

The spin wave propagation route depends on the mutual arrangement of magnets in the magnonic mesh. This is the keystone of MCM operation. In order to illustrate it, we present the results of numerical modeling. In Fig.4, there is shown an equivalent circuit for MCM. The mesh of waveguides with magnets is replaced with a 2D mesh of impedances. The real and the imaginary part of each cell corresponds to the spin wave attenuation and phase shift accumulated during propagation through the junction. Rigorously speaking, there should be additional cells in the mesh (i.e., impedances) to account for spin wave damping and accumulated phase shift during the propagation between the junctions. To simplify our consideration, we assume that the damping and the phase shift during the propagation between the junctions are much smaller compared to the ones during the propagation through the junction. The real and the imaginary parts of the junction impedances are frequency dependent. They may be obtained from micromagnetic simulations [20] or from experiment [21].

There are three steps in the modeling procedure. First, one needs to find the total attenuation and the phase shift produced by the passive part *for all possible frequencies*.  Second, the obtained results are checked to find the frequency(s) at which the self-oscillation conditions (1.1) and (1.2) are met. Finally, one needs to find the map of spin wave power flow through the mesh at the resonant frequency(s). The most time-consuming is the first step as it takes a number of subsequent calculations to find the mesh responses in a wide frequency range. In order to speed up calculations and illuminate the essence of the proposed memory, we make several assumptions. (i) We assume that each propagation route in the mesh is associated with a certain propagation frequency. (ii) The attenuation is linearly proportional to the propagation distance.  (iii) The junctions provide a frequency-independent phase shift. The objective is to show the change in the signal propagation route depending on the arrangement of a given set of elements in the mesh.

In Fig.5(a), it is shown a mesh where numbers in the boxes show the phase shift accumulated by the propagating wave. It is assumed that the output attenuators a set to exclude the routes for more than six junctions (i.e., the amplitude condition is satisfied for all routes coming through five or six junctions). The output phase shifters are set to $\Psi = 0.5\pi$. The phase condition is satisfied for the routes providing $1.5\pi$ phase shift. The particular frequency(s) coming through these routes are of no importance.  All the input and output switches are in the On state. The set of power detectors in Fig.5(a) show the signal propagation route for the given configuration of magnets. There are two routes. One route is from the input port #2 to the output port #1. The other route is from input port #5 to the output port #4. Let us change the places of two cells in the mesh. For example, we flip the two adjacent cells in the second row at columns two and three.  It results in the change of the signal propagation route. In Fig.5(b), the green circles of power detectors show the routes. It is different from the one in Fig.5(a). The difference in the memory state is eight bits. It is possible to engineer magnonic meshes where the change in the position of just one magnet will lead to the tens-bit difference in the output.

It should be noted that changing the mutual position of cells may or may not change the propagation routes in the mesh. Fig.6(a) shows the propagation route (i.e., green circles) for the arrangement of cells as in Fig.5(a) but for the output phase shifter set for $\Psi = 0.6\pi$. The phase condition (1.2) is satisfied for the routes that provide $1.4\pi$ phase shift. As in the previous example, the output attenuators a set to exclude the routes through more than six junctions. There is just one route from input port # 2 to output port #3 that provides the required phase shift. Flipping the cells (i.e., second row, second and third columns) does not change the propagation route. It is illustrated in Fig.6(b). The examples shown in Figs. 5 and 6 are aimed to illustrate the plethora of combinations in the mutual cell positions leading or not leading to the different outputs. The freedom of engineering the output by changing the position of cells in the mesh is important for MCM programming.

### IV. Experimental data

In this part, we present experimental data obtained for the prototype with just three magnets. The schematics of the prototype are shown in Fig.7(A). It is a multi-path active ring circuit comprising electric and magnetic parts. The electric part consists of an amplifier (three amplifiers Mini-Circuits, model ZX60-83LN-S+ connected in series), and a phase shifter (ARRA, model 9418A). The magnonic part is a ferrite film with three fixed places (i.e., depicted by the circles numbered 1,2, and 3) for three different magnets to be placed on top of the film. There are three input and three output antennas to connect electric and magnetic parts. Each input/output port can be independently connected/disconnected. There are three voltage-tunable bandpass filters at the output ports. The filters are commercially available YIG-based frequency filters produced by Micro Lambda Wireless, Inc, model MLFD-40540. The experimental data on the filter transmission and phase delay can be found in the supplementary materials. The filters at output ports #1, #2, and #3 are set to the central frequencies $f_1 = 2.539$ GHz, $f_2 = 2.475$ GHz and $f_3 = 2.590$ GHz, respectively. Magnonic and electric parts are connected via the set of splitters and combiners (i.e., SPLT 1-3, Sigatek SP11R2F 1527). The power at each output port and the total power circulating in the ring circuit is measured by the spectrum analyzer (SA) GW Instek GSP-827 connected to the circuit through a directional coupler (DC, KRYTAR, model 1850). There are four places of connection shown in Fig.7(A). $P_0$ is the total power in the circuit measured just after the amplifier, $P_1$, $P_2$, and $P_3$ are the powers measured at the output ports after signal propagation through the passive magnonic part. The signal is significantly damped after the passive part so the sum of $P_1, P_2$, and $P_3$ is not equal to $P_0$. SA is also used for detecting the frequencies of the auto-oscillations in the ring circuit. A more detailed schematics of the experimental setup with a detailed map of signal attenuation through the parts can be found in the Supplementary materials.

The cross-section of the passive magnonic part is shown in Fig.7(B). It consists from the bottom to the top from a permanent magnet made of NdFeB, a Printed Circuit Board (PCB) substrate with six short-circuited antennas, a ferrite film made of GGG substrate and YIG layer, and a plastic plate with three pits for magnets to be inserted. The permanent magnet is aimed to create a constant bias magnetic field. Hereafter, we refer to this relatively big permanent magnet (model BX8X84 by K&J Magnets, Inc., dimensions 1.5"x1.5"x0.25") as a magnet in the text. The magnetic field produced by this magnet defines the frequency window as well as the type of spin waves that can propagate in the ferrite film. The bias field is about 375 Oe and directed in-plane on the film surface. The photo of the PCB with six antennas is shown in Fig.7(c). The antennas are marked as 1,2,3,..6. The characteristic size of the antenna is 2 mm length and 0.15 mm width. The ferrite film is made of YIG grown by liquid epitaxy on GGG substrate. YIG was chosen due to the low spin wave damping. The film is not patterned. The thickness of the film is 42 µm. The saturation magnetization is close to 1750 G, the dissipation parameter (i.e., the half-width of the ferromagnetic resonance) ΔH = 0.6 Oe. The plastic plate is mechanically attached on top of the ferrite film. There are three pits drilled in the layer for placing the micro-magnets. There are three NdFeB micro-magnets of volumes 0.02 mm$^3$, 0.03 mm$^3$, and 0.06 mm$^3$, respectively. The magnets are placed inside plastic tubes of different color. The smallest-volume magnet is placed into the tube of black color without a sticker. The tubes with the white and the red stickers correspond to the middle-volume and large-volume magnets respectively. Hereafter, we refer to the micromagnets as Black (B), White (W), and Red (R). The photo of the devices with tubes can be found in the Supplementary materials.

The first set of experiments was accomplished for the case with three input and three output antennas connected. Antennas marked as # 3, #4, and #6 are used for spin wave excitation in the ferrite film. Antennas marked as #1, #2, and #5 are used for detecting the inductive voltage produced by the spin waves at the output. The summary of experimental data are shown in Table 1. The first column shows the combination of input and output switches. For instance, (111) means that all three input antennas are connected to the electric part. The second column shows the position of the external phase shifter. The external phase is set to $\Psi = 0\pi$. The third column shows the magnet arrangement. For example, BWR means that B magnet (smallest) is inserted into the pit #1, W magnet (medium) is inserted in the pit #2, and R magnet (largest) is inserted into the pit #3. Combination (000) stands for the case without magnets placed in the pits. The fourth column shows the frequencies of the auto-oscillations (i.e., measured by SA). It may be one or several frequencies at the same time. The fifth column shows the power of the auto-oscillation $P_0$ at different frequencies. For example, the numbers in the second row (+2dBm and +4 dBm) correspond to the frequencies 2.590 GHz, and 2538 GHz, respectively. The sixth column shows the power measured at the three output ports, where three numbers in each row correspond to $P_1$, $P_2$, and $P_3$, respectively. The output power ranges from -30 dBm to -79 dBm. The last column in the table shows the logic output. It is logic 1 if the output power exceeds -45 dBm and logic 0, otherwise. Power below the reference value is shown in blue color, while the power larger than reference power is shown in red color. For example, -27 dBm, -75 dBm, -73 dBm in the first row correspond to the memory state 100. The data presented in Table 1 provides a detailed picture of the active ring dynamics including the frequencies of auto-oscillation, the distribution of power between the frequencies, and the analog output at each port. There is no need in using SA in a practical device. All the collected data is aimed to explain the physical origin of data storage in MCM. The memory device will only provide binary output for the given binary address.

In Tables 2, there are shown experimental data for the case with three input and three output antennas connected. The external phase is set to $\Psi = 0.63\,\pi$. One can see the difference in the output power distribution compared to the previous case. There three frequencies of auto-oscillation for BWR magnet configuration. In Table 3, there are shown experimental data for the case with three input and three output antennas connected. The external phase is set to $\Psi = 1.25\,\pi$. In Table 4, there are shown experimental data for the case with three input and three output antennas connected. The external phase is set to $\Psi = 1.75\,\pi$. The collected data reveal the complex dynamic of auto-oscillations in the electro-magnetic active ring circuit. There may be one, two, or three different frequencies of auto-oscillations. It provides all possible frequency combination at the output resulting in the different output state. There is one just output state (000) missing in the collected data. This state appears when the amplitude condition (1.1) is not satisfied. We intentionally kept low attenuation and relatively high amplification to demonstrate the dependance output dependence on the external phase.

The absence of magnet in the pit can be considered as an additional state. In Table 5, there are shown experimental data for the case when pit #1 is empty. The other two pits were used for different arrangement of magnets. The external phase is set to $\Psi = 0\,\pi$. There is only one frequency of auto-oscillation observed for all arrangements of magnets. It may give an insight on the importance of information stored in one magnet on top of ferrite film (e.g., frequency-selective attenuation and phase shift) that define spin wave routing.

Finally, the system was studied for a different combination of input/output switches. In Table 6, there are shown experimental data for the case with two input (#3 and #4) and three output antennas connected. The external phase is set to $\Psi = 0\,\pi$. The frequencies of auto-oscillation, frequency combination, and the output states are different compared to the case with three input ports. The addition of one more input lead to the disappearance of some frequencies. For example, one can compare the second and the fourth rows in Table 1 and Table 5. The disappearance of auto-oscillation on some frequencies can be well explained by the interference (e.g., destructive interference) of spin waves coming from different inputs. In any rate, the addition of extra input/output channel cannot be described as a simple superposition of routes.

The data shown in Tables 1-5 are obtained in the active ring configuration. It takes less than a millisecond for the system to reach the self-sustained oscillations. In general, the time required to reach the steady-state regime depends on many factors including spin wave dispersion, parameters of electric broadband amplifier, etc. Raw data on spin wave transport in the circuit can be found in the Supplementary materials.

## V. Discussion

There are two important observations we want to make based on the obtained experimental data. (i) Spin wave propagation route(s) does depend on the configuration of magnets on top of ferrite waveguide, the combination of input and output ports, and the output phase shifter. For instance, the arrangement of three different magnets or arrangement of two different magnets with one empty pit results in the different spin wave propagation routes. The reason for spin wave re-routing is the difference in the magnetic field profile on the top of ferrite film that appears for different arrangement of magnets. The re-routing can be modeled considering magnet/ferrite film as a bandpass filter and a phase shifter as illustrated in Section 3. However, the high-fidelity numerical modeling would require an enormous deal of work to link the magnetic film profile to spin wave propagation routes in a wide frequency range. External phase is an additional parameter which affects spin wave propagation in the active ring circuit. It makes a fundamental difference with conventional RAM where low/high electric current is directly related on the high/low resistance states of the memory cells. In turn, the phase-dependent transport allows us exploit the different combination of input/output ports. Also, the addition of extra ports is not equivalent to adding additional routes. It may happen that some frequencies (routes) disappear for a larger number of inputs due to the spin wave interference.

(ii) Spin wave propagation routes can be recognized by the set of power sensors with high accuracy. In the presented experiments, spin wave power was measured only at the output ports (i.e., no sensors within the matrix). The On/Off ratio (i.e., the difference between the outputs where most of power flows and the outputs with minimum power) exceeds 35 dB at room temperature. It makes it possible to tolerate the inevitable structure imperfections, the difference in the efficiency of input/output antennas, etc. This big ratio is achieved by the introduction of frequency filters aimed to separate frequency response between the different outputs. It may be possible to achieve even bigger On/Off ration by using filters with a smaller bandwidth. It will take an additional comparator-based circuit to digitize MCM output.

These observations confirm the main idea of this work on the feasibility of data storage using spin wave propagation routes. It inherent the advantages of traditional magnetic-based memory including non-volatility and a long retention time. At the same time, MCM provides a fundamental advantage in data storage density compared to the existing memory devices. To comprehend this advantage, we extracted the data from Tables 1-4 obtained for the same set of external phase shifter but with the different configuration of magnets. In Fig.7, there are shown the arrangement of the magnets on the left and corresponding truth tables for four phases on the right. There are three bits corresponding to the memory state in each row. Conventional memory with three magnets stores just three bits. It is the same amount of data that can be read-out at one given phase in MCM (e.g., phase = $0\,\pi$). All other rows (i.e., 3 out of 4 in each table) contain an extra or exceed information compared to conventional RAMs. According to Eq.(3), the advantage over the conventional data storage devices scales according to the power law with the size of magnonic mesh $n$.

There are several critical comments to mention. (i) The structure of the prototype is different from the general view MCM shown in Fig.3(a). There are only three output ports in the prototype. The addition of power sensors within the magnetic part (e.g., between the magnets) would give a better picture of spin wave power distribution. It is also not clear if the ISHE sensors would need frequency filters for better spin wave route detection. (ii) The read-in procedure is not described. The need to use magnets with more than two thermally stable states of magnetization (e.g., magnets of different sizes) significantly complicated the fabrication and initialization procedure. Theoretically, it can be achieved during the fabrication or using specially engineered micro magnets with after-fabrication initialization. We want also to refer to the recently reported nanomagnet reversal by propagating spin waves [22]. The switching was observed in ferromagnet/ferrimagnet hybrid structures consisting of $Ni_{81}Fe_{19}$ nanostripes prepared on top of YIG film. It may be a convenient way for MCM programming.

The key question regarding MCM potential advantages in data storage is associated with the possibility of programming. *Is it possible to find a magnet arrangement for any given truth table?* It is an NP-hard problem to check all possible magnet configurations with the hope to find the desired one. On the other hand, it may be possible to utilize only a fraction of input-output correlations for data storage. This and many other questions and concerns deserve special consideration combining approaches developed in combinatorics, physics, and magnonics. This work is aimed to introduce the concept of MCM and outline its potential advantages.

## VI. Conclusions

We described a novel type of magnetic memory which is aimed to exploit the mutual arrangement of magnets for data storage. The principle of operation is based on the correlation between the arrangement of magnets on top of ferrite film and the spin wave propagation routes. The number of routes scales factorial with the number of magnets that makes it possible to encode more information compared to conventional magnetic memory devices exploiting the individual states of magnets. We presented experimental data on the proof-of-the-concept experiments on the prototype with just three magnets placed on top of a of single-crystal yttrium iron garnet $Y_3Fe_2(FeO_4)_3$ (YIG) film. The results demonstrate a robust operation with an On/Off ratio for route detection exceeding 35 dB at room temperature. This work is a first step toward the novel type of combinatorial memory devices which have not been explored. The material structure and principle of operation of MCM are much more complicated compared to conventional RAMs. At the same time, MCM may pave the road to unprecedented data storage capacity where a device with just 25 magnets can store as much as $25! = 10^{25}$ bits.

**Competing financial interests**

The authors declare no competing financial interests.

**Data availability**

All data generated or analyzed during this study are included in this published article.

**Acknowledgment**

This work of M. Balinskiy and A. Khitun was supported in part by the INTEL CORPORATION, under Award #008635, Project director Dr. D. E. Nikonov, and by the National Science Foundation (NSF) under Award # 2006290, Program Officer Dr. S. Basu. The authors would like to thank Dr. D. E. Nikonov for the valuable discussions.

**Figure Captions**

Figure 1. (A) Schematics of RAM organization. There is a 2D mesh of memory cells arranged in cows and columns. The addressing of a required cell is accomplished by the row/column decoders. (B) Example of RAM truth table. The first column in the table depicts the memory cell binary address. The second column in the table depicts the cell state. The maximum number of bits stored is limited by the number of memory cells.

Figure 2. (A) Schematics of a two-path active ring circuit. The active part includes a nonlinear broadband amplifier $G$, a phase shifter $\Psi$, and an attenuator $A$. The passive part consists of two paths. Each path has a bandpass frequency filter, a phase shifter, and a power sensor $P$. The bandpass filters on the top and the bottom paths are set to the different frequencies $f_1$ and $f_2$, respectively. The phase shifters provide tifferent phase shifts $\Delta_1 = 1.5\pi$ and $\Delta_2 = 0.5\pi$. (B) The auto oscillation power as a function of the external phase shifter $\Psi$. The auto oscillations occur on the frequencies $f_1$ and $f_2$ when the phase condition (1.2) is met. The insets illustrate the signal propagation route. The green circle depicts the power flow.

Figure 3. (A) Schematics of MCM. The core of the structure is a mesh of magnonic waveguides shown in the light blue color. There are magnets depicted by the blue-red rhombs placed on top of waveguide junctions. There are three different sizes for the magnets, where each magnet has eight thermally stable states of magnetization. There are power sensors placed on top of the waveguides between the junctions. The mesh is included into an active ring circuit via switches on the left and right sides. There are voltage-tunable frequency filters $f_i$, voltage-tunable phase shifters $\Psi_i$, and voltage-tunable attenuators $A_i$ at each output port (right side). There is one broadband amplifier $G$ in the electric part. (B) Memory truth table. The left column shows the memory address. It includes the binary number corresponding to sthe tates the input switches, the binary number corresponding to the output switches, and the binary number corresponding to the states of the phase shifters $\Psi_i$. The right column shows the memory state, which is set of digitized sensor outputs.

Figure 4. Equivalent circuit for MCM. The mesh of waveguides with magnets is replaced with a mesh of cells with impedances $Z(f)_{ij}$, where subscripts $i$ and $j$ correspond to the column and row numbers, respectively. The real and the imaginary part of each cell are frequency dependent and correspond to the spin wave attenuation and phase shift accumulated while propagating through the junction. The output bandpass filters, phase shifters, and attenuators are replaced by the frequency depend impedances $Z(f)_j$, where subscript $j$ corresponds to the row number.

Figure 5. Schematics of the simplified circuit model. The numbers in the boxes show the phase shift in the cell. The amplitude condition is satisfied for all routes coming through five or six junctions. The output phase shifters are set to $\Psi = 0.5\pi$. All input and the output switches are in the On state. (A) The set of power sensors (green circles) shows the signal propagation routes for the given configuration of phase shifters. (B) The set of power sensors (green circles) shows the signal propagation routes where two cells are flipped in their positions (two adjacent cells in the second row at columns two and three).

Figure 6. Schematics of the simplified circuit model. The numbers in the boxes show the phase shift in the cell. The amplitude condition is satisfied for all routes coming through five or six junctions. The output phase shifters are set to $\Psi = 0.6\pi$. All input and the output switches are in the On state. (A) The set of power sensors (green circles) shows the signal propagation routes for the given configuration of phase

shifters. (B) The set of power sensors (green circles) shows the signal propagation routes where two cells are flipped in their positions (two adjacent cells in the second row at columns two and three).

Table 1. Experimental data obtained for different configurations of magnets on top of ferrite film. All three input and output antennas are operating. The external phase is set to $\Psi = 0\pi$.

Table 2. Experimental data obtained for different configuration of magnets on top of ferrite film. All three input and output antennas are operating. The external phase is set to $\Psi = 0.63\,\pi$.

Table 3. Experimental data obtained for different configuration of magnets on top of ferrite film. All three input and output antennas are operating. The external phase is set to $\Psi = 1.25\,\pi$.

Table 4. Experimental data obtained for different configuration of magnets on top of ferrite film. All three input and output antennas are operating. The external phase is set to $\Psi = 1.75\,\pi$.

Table 5. Experimental data obtained for different configuration of magnets on top of ferrite film where pit #1 is empty. The other two pits are used for the different arrangements of two magnets. All three input and output antennas are operating. The external phase is set to $\Psi = 0\,\pi$.

Table 6. Experimental data obtained for different configuration of magnets on top of ferrite film. Three are two input and three output antennas connected. The external phase is set to $\Psi = 0\pi$.

Table 7. Summary of experimental data from Tables 1-4. The mutual arrangement of magnets is shown on the left. The tables on the right show the output of MCM.

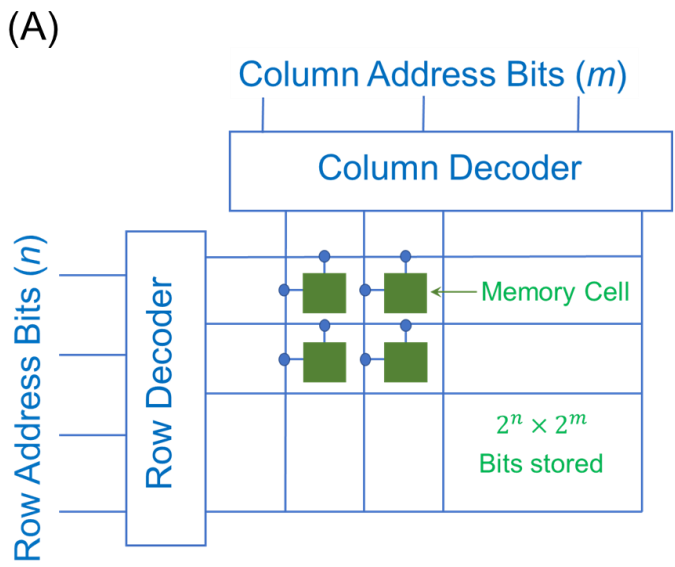

**Figure 1**

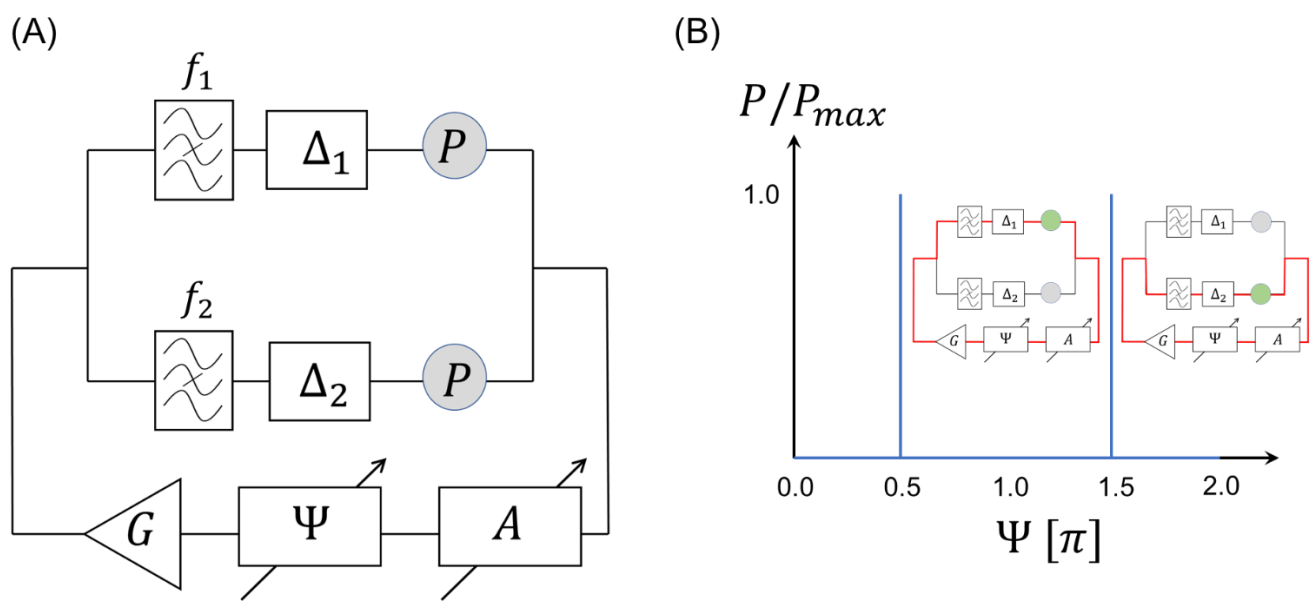

**Figure 2**

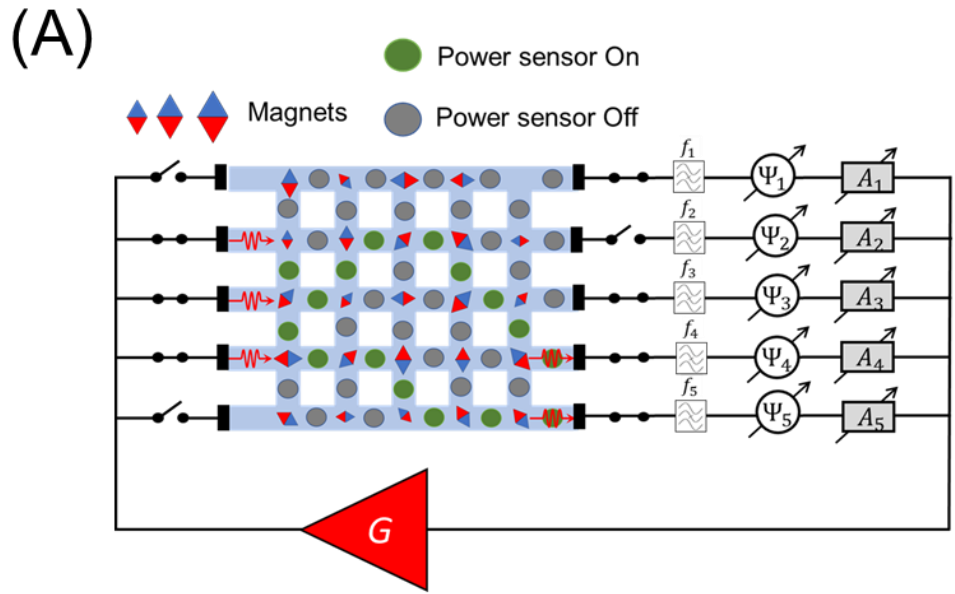

| Address | | | State |
|---|---|---|---|
| Input switches | Output switches | Output Phases | Sensor Outputs |
| 00001 | 00001 | 000 000 000 000 000 | 11011 |
| | | | 01110 |
| | | | 00000 |
| | | | 00000 |
| | | | 00000 |
| | | | 00000 |
| | | | 00000 |
| | | | 00000 |
| | | | 00000 |
| ⋮ | | | |
| 01110 | 10111 | 000 000 000 000 000 | 00000 |
| | | | 00000 |
| | | | 01100 |
| | | | 11010 |
| | | | 10010 |
| | | | 10001 |
| | | | 11001 |
| | | | 00100 |
| | | | 00111 |
| ⋮ | | | |
| 11111 | 11111 | 111 111 111 111 111 | 11111 |
| | | | 00000 |
| | | | 11111 |
| | | | 11111 |
| | | | 0000 |
| | | | 10001 |
| | | | 11111 |
| | | | 11111 |
| | | | 00111 |

**Figure 3**

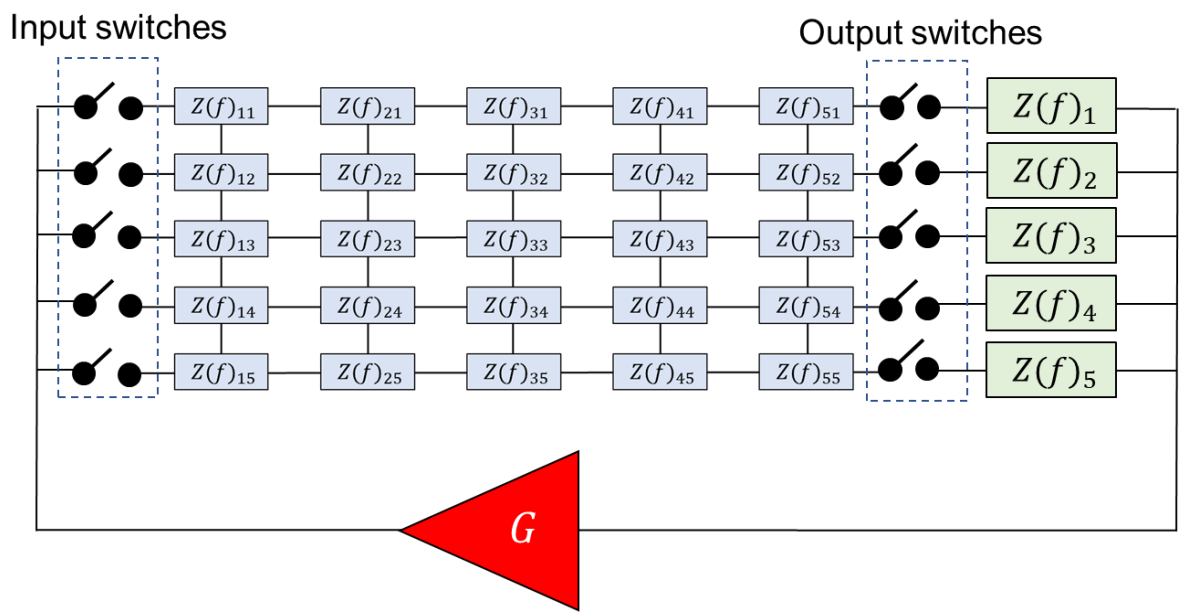

**Figure 4**

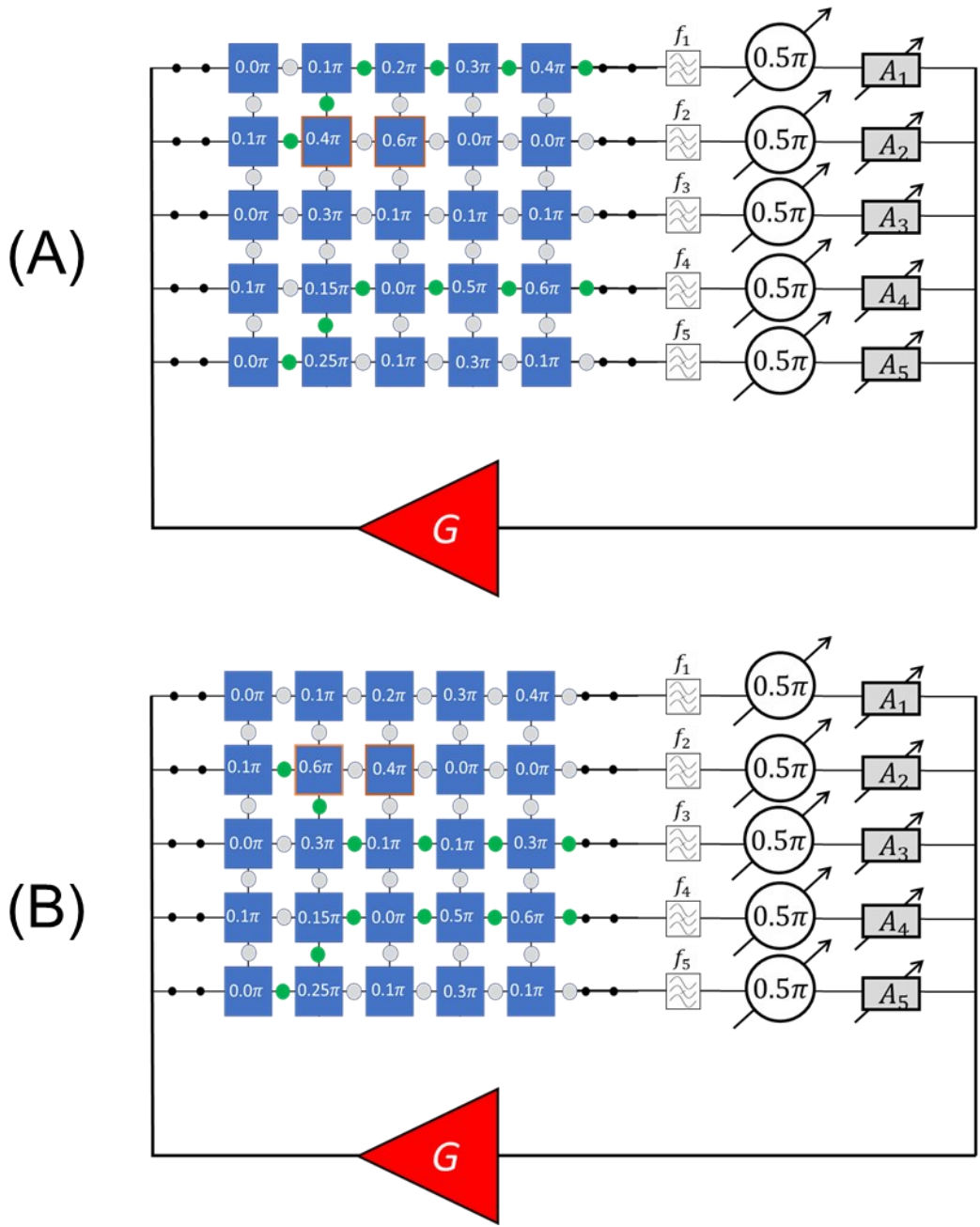

**Figure 5**

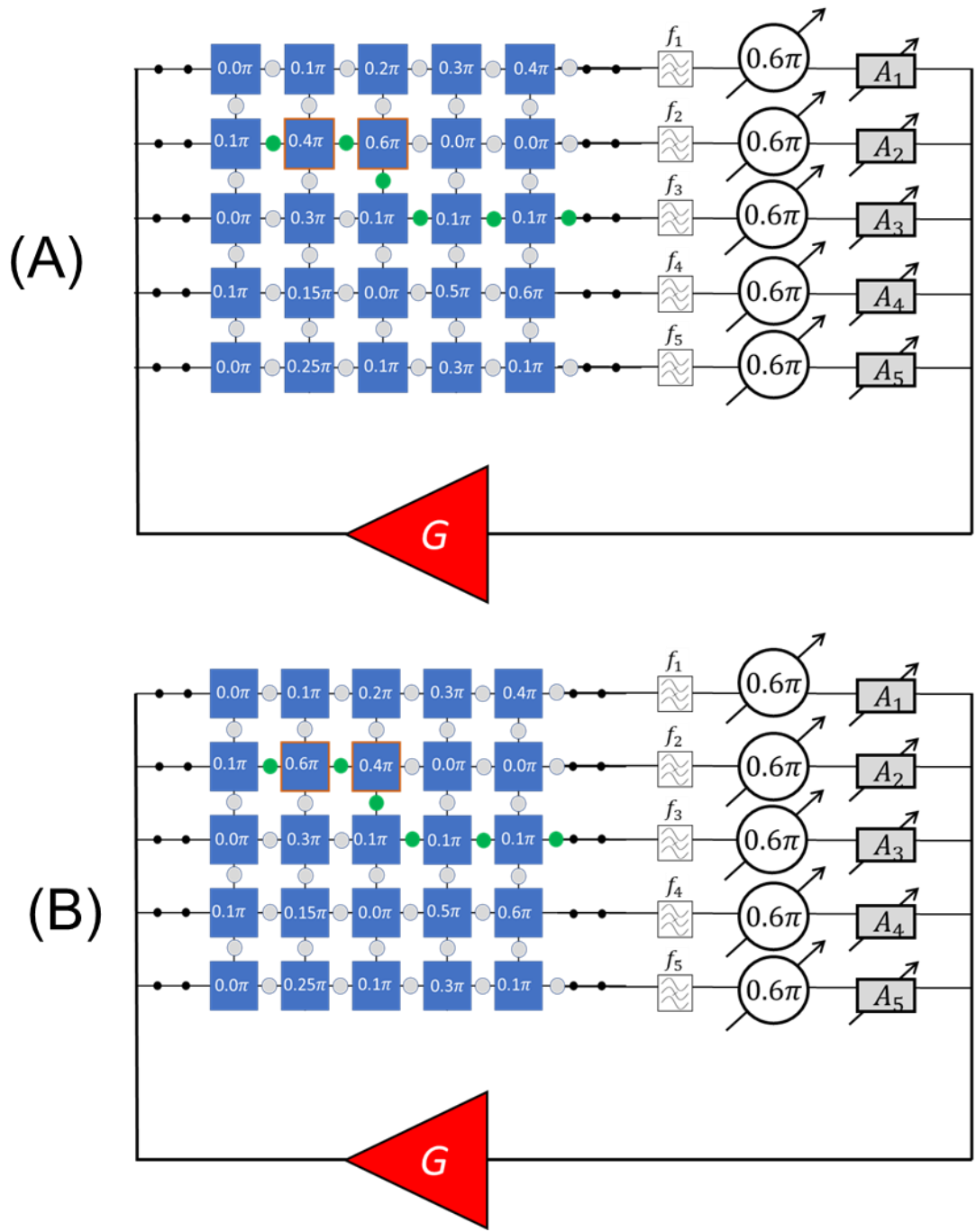

**Figure 6**

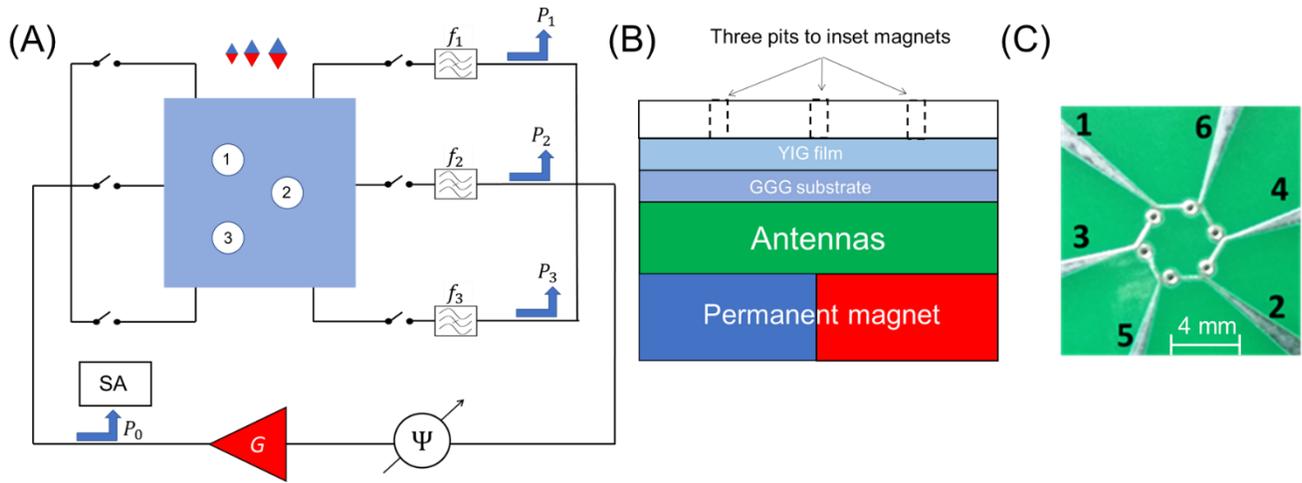

**Figure 7**

| Switches Input | Switches Output | Phase [π] | Magnets Arrangement | Auto-oscillation frequency [GHz] | $P_0(f)$ [dBm] | $P_1, P_2, P_3$ [dBm] | Memory State |
|---|---|---|---|---|---|---|---|
| 111 | 111 | 0 | 000 | 2.590 | +7.5 | -27, -75, -73 | 100 |
| 111 | 111 | 0 | BWR | 2.590, 2.538 | +2, +4 | -30, -78, -27 | 101 |
| 111 | 111 | 0 | BRW | 2.475 | +6 | -78, -30, -71 | 010 |
| 111 | 111 | 0 | WBR | 2.539 | -1 | -72, -71, -39 | 001 |
| 111 | 111 | 0 | WRB | 2.539 | +2 | -75, -69, -37 | 001 |
| 111 | 111 | 0 | RBW | 2.590, 2.539 | +3, +1 | -39, -75, -37 | 101 |
| 111 | 111 | 0 | RWB | 2.539 | -3 | -81, -78, -36 | 001 |

**Table 1**

| Switches Input | Output | Phase [π] | Magnets Arrangement | Auto-oscillation frequency [GHz] | $P_0(f)$ [dBm] | $P_1, P_2, P_3$ [dBm] | Memory State |
|---|---|---|---|---|---|---|---|
| 111 | 111 | 0.63 | 000 | 2.590 | +6 | -31, -70, -68 | 100 |
| 111 | 111 | 0.63 | BWR | 2.590, 2.475, 2538 | +2, +4, -1 | -33, -28, -39 | 111 |
| 111 | 111 | 0.63 | BRW | 2.590, 2.475 | +6, 0 | -27, -31, -71 | 110 |
| 111 | 111 | 0.63 | WBR | 2.539, 2.539 | -3, -1 | -36, -73, -36 | 101 |
| 111 | 111 | 0.63 | WRB | 2.539 | +4 | -77, -73, -39 | 001 |
| 111 | 111 | 0.63 | RBW | 2.590, 2.475 | +5, +1 | -37, -39, -78 | 110 |
| 111 | 111 | 0.63 | RWB | 2.475, 2.539 | 0, +2 | -79, -36, -33 | 011 |

**Table 2**

| Switches Input Output | Phase [π] | Magnets Arrangement | Auto-oscillation frequency [GHz] | $P_0(f)$ [dBm] | $P_1, P_2, P_3$ [dBm] | Memory State |
|---|---|---|---|---|---|---|
| 111   111 | 1.25 | 000 | 2.590 | +6 | -31, -70, -68 | 100 |
| 111   111 | 1.25 | BWR | 2.475 | +3 | -73, -31, -71 | 010 |
| 111   111 | 1.25 | BRW | 2.590, 2.475 | +5, +1 | -27, -31, -71 | 110 |
| 111   111 | 1.25 | WBR | 2.475, 2.539 | 0, +3 | -79, -35, -31 | 011 |
| 111   111 | 1.25 | WRB | 2.590, 2.475 | +1, +2 | -35, -34, -75 | 110 |
| 111   111 | 1.25 | RBW | 2.590, 2.539 | +3, +3 | -30, -78, -30 | 101 |
| 111   111 | 1.25 | RWB | 2.590, 2.475, 2.539 | +2, +3, -1 | -33, -30, -36 | 111 |

**Table 3**

| Switches Input Output | Phase [π] | Magnets Arrangement | Auto-oscillation frequency [GHz] | $P_0(f)$ [dBm] | $P_1, P_2, P_3$ [dBm] | Memory State |
|---|---|---|---|---|---|---|
| 111  111 | 1.75 | 000 | 2.590 | +5 | -30, -70, -69 | 100 |
| 111  111 | 1.75 | BWR | 2.539 | +3 | -75, -73, -35 | 001 |
| 111  111 | 1.75 | BRW | 2.475, 2.539 | 0, +3 | -79, -35, -31 | 011 |
| 111  111 | 1.75 | WBR | 2.539 | +4 | -77, -73, -36 | 011 |
| 111  111 | 1.75 | WRB | 2.475, 2.539 | +3, +3 | -78, -33, -39 | 011 |
| 111  111 | 1.75 | RBW | 2.590, 2.475, 2.539 | +2, +3, -1 | -33, -30, -36 | 111 |
| 111  111 | 1.75 | RWB | 2.475, 2.539 | +2, +4 | -75, -36, -33 | 011 |

**Table 4**

| Switches Input Output | Phase [π] | Magnets Arrangement | Auto-oscillation frequency [GHz] | $P_0(f)$ [dBm] | $P_1, P_2, P_3$ [dBm] | Memory State |
|---|---|---|---|---|---|---|
| 111  111 | 0 | 000 | 2.579 | +6 | -33, -81, -79 | 100 |
| 111  111 | 0 | 0WR | 2.466 | +3 | -80, -33, -73 | 010 |
| 111  111 | 0 | 0RW | 2.466 | +5 | -79, -33, -78 | 010 |
| 111  111 | 0 | 0BW | 2.579 | +3 | -31, -81, -75 | 100 |
| 111  111 | 0 | 0WB | 2.466 | +2 | -81, -33, -75 | 010 |
| 111  111 | 0 | 0BR | 2.466 | 0 | -78, -33, -79 | 010 |
| 111  111 | 0 | 0RB | 2.523 | +3 | -69, -72, -27 | 001 |

**Table 5**

| Switches Input Output | Phase [π] | Magnets Arrangement | Auto-oscillation frequency [GHz] | $P_0(f)$ [dBm] | $P_1, P_2, P_3$ [dBm] | Memory State |
|---|---|---|---|---|---|---|
| 110   111 | 0 | 000 | 2.631 | +3 | -30, -81, -78 | 100 |
| 110   111 | 0 | BWR | 2.631, 2.492, 2.522 | +2, +2, +2 | -32, -33, -34 | 111 |
| 110   111 | 0 | BRW | 2.631 | +6 | -32, -84, -79 | 100 |
| 110   111 | 0 | RBW | 2.631, 2.522 | +4, +4 | -31, -79, -34 | 101 |
| 110   111 | 0 | RWB | 2.631 | +2 | -33, -81, -75 | 100 |
| 110   111 | 0 | RBW | 2.522 | +2 | -81, -84, -30 | 001 |
| 110   111 | 0 | WRB | 2.631, 2.492 | +3, +3 | -35, -33, -79 | 110 |

**Table 6**

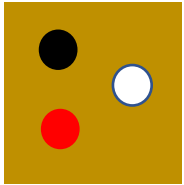

BWR

| Input Phase [π] | Output State |
|---|---|
| 0 | 101 |
| 0.63 | 111 |
| 1.25 | 010 |
| 1.75 | 001 |

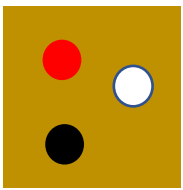

RWB

| Input Phase [π] | Output State |
|---|---|
| 0 | 001 |
| 0.63 | 011 |
| 1.25 | 110 |
| 1.75 | 011 |

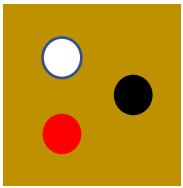

WBR

| Input Phase [π] | Output State |
|---|---|
| 0 | 001 |
| 0.63 | 101 |
| 1.25 | 011 |
| 1.75 | 001 |

**Table 7**

Supplementary Materials for

**Magnonic Combinatorial Memory**

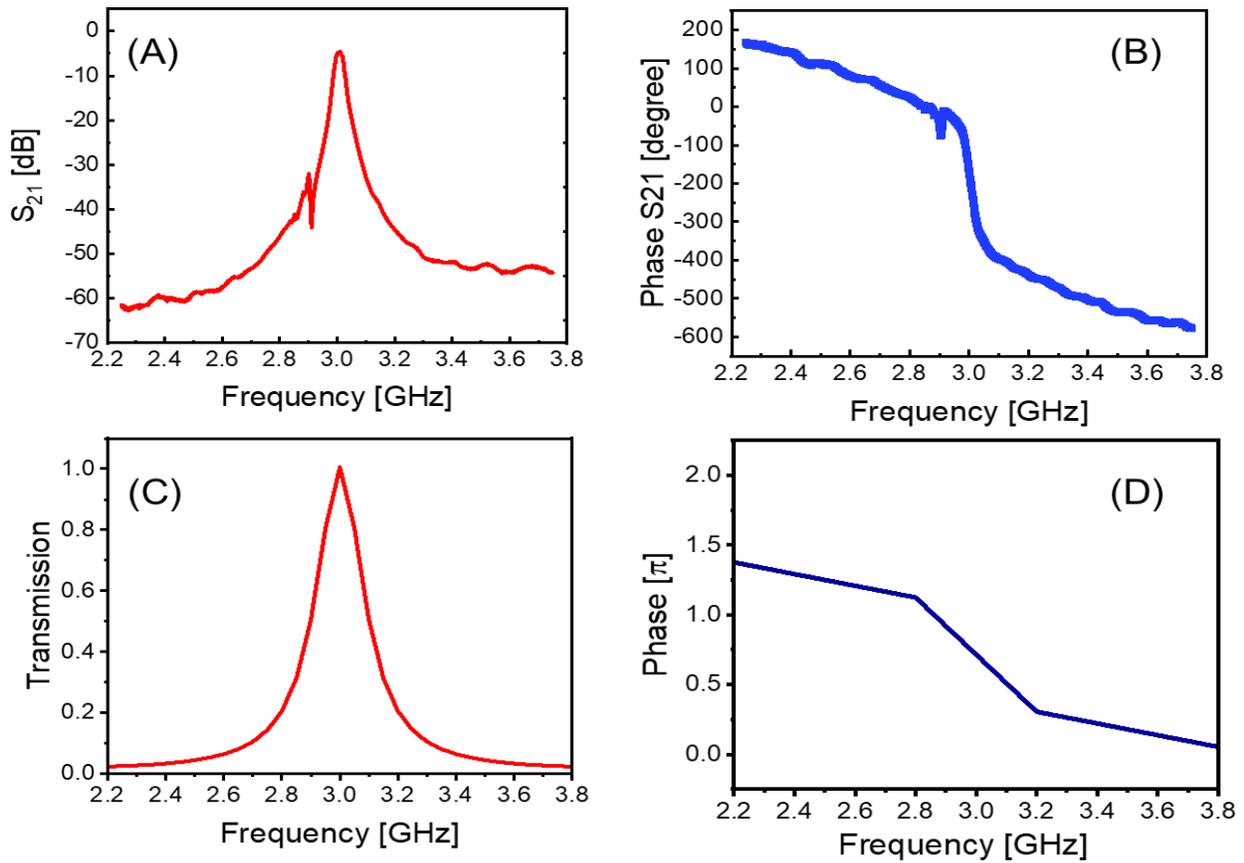

Figure [s1]: YIG-based frequency filter produced by Micro Lambda Wireless, Inc, model MLFD-40540. (A) Experimental data: S21 parameter (amplitude) of the commercial filter. (B) Experimental data: S21 parameter (phase shift) of the commercial filter. (C) Results of numerical fitting: transmission of the spin wave element. (D) Results of numerical fitting: phase shift produced by the spin wave element. The data are shown in the frequency range from 2.2 GHz to 3.8 GHz.

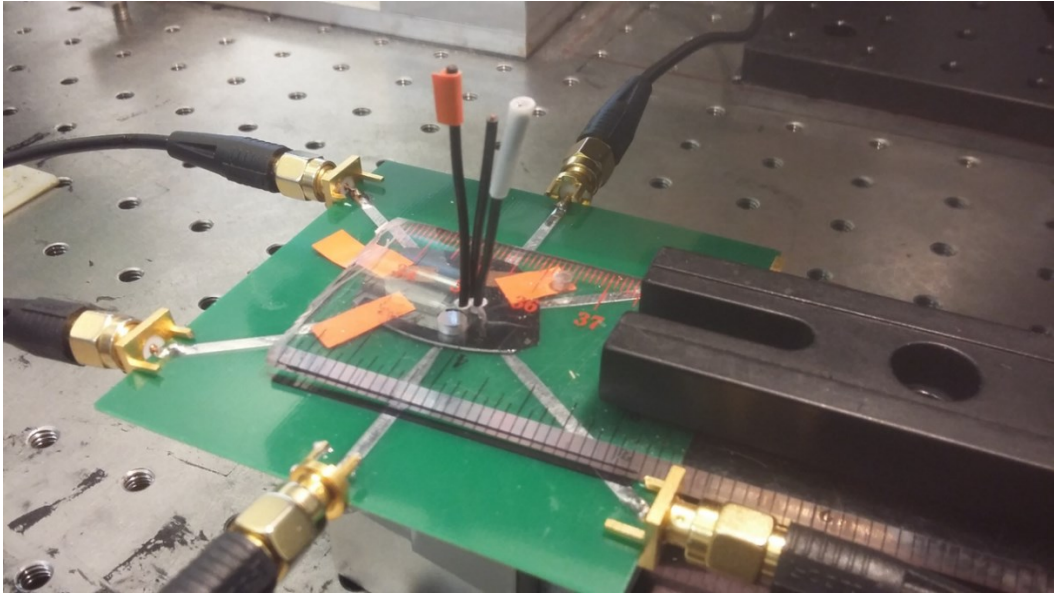

Figure [s2]: Photo of the passive part with six antennas connected to the electric part via coaxial cables. There are seen three tubes labeled as Red, Black, and White. There are three NdFeB micro-magnets of volumes 0.02 mm$^3$, 0.03 mm$^3$, and 0.06 mm$^3$, placed in the Black (B), White (W), and Red (R) tubes, respectively.

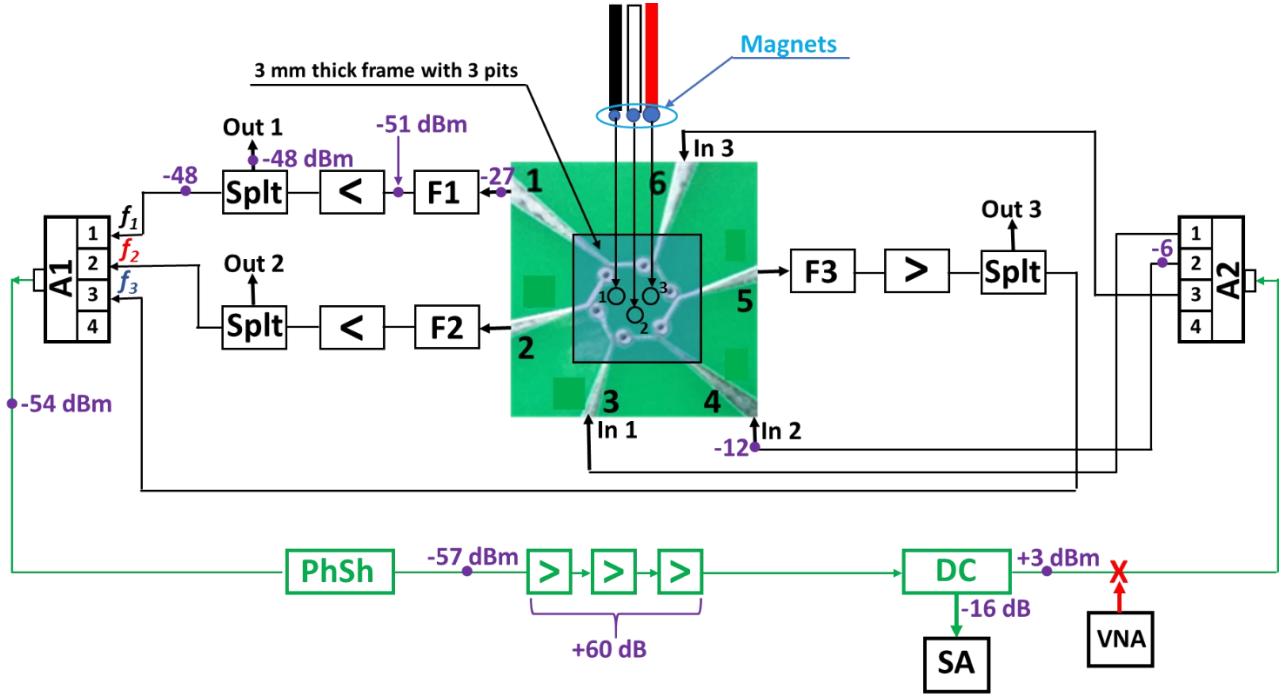

Figure [s3]: Schematics of the experimental setup. The passive part with three tubes with magnet is shown in the center. There are six antennas (e.g., three input and three output ports) connecting the passive and the active parts. Magnonic and electric parts are connected via the set of splitters and combiners (i.e., SPLT 1-3, Sigatek SP11R2F 1527). Each of the output ports is equipped with a frequency filter (F1, F2, F3) and an amplifier (Mini-Circuits, model ZX60-83LN-S+) that is used for initial calibration. There is a prominent signal attenuation introduced by the different parts in the circuit (e.g., the device, splitters, combiners, frequency filters, etc.). To comprehend the level of signal attenuation/amplification, there are shown numbers in pink colors. These numbers indicate the power level in the corresponding points of the schematic. There is also a programmable network analyzer (PNA) Keysight Technologies, model N5221A-217 for measuring S21 parameter of the circuit. The PNA is included in the circuit by breaking the active ring configuration.

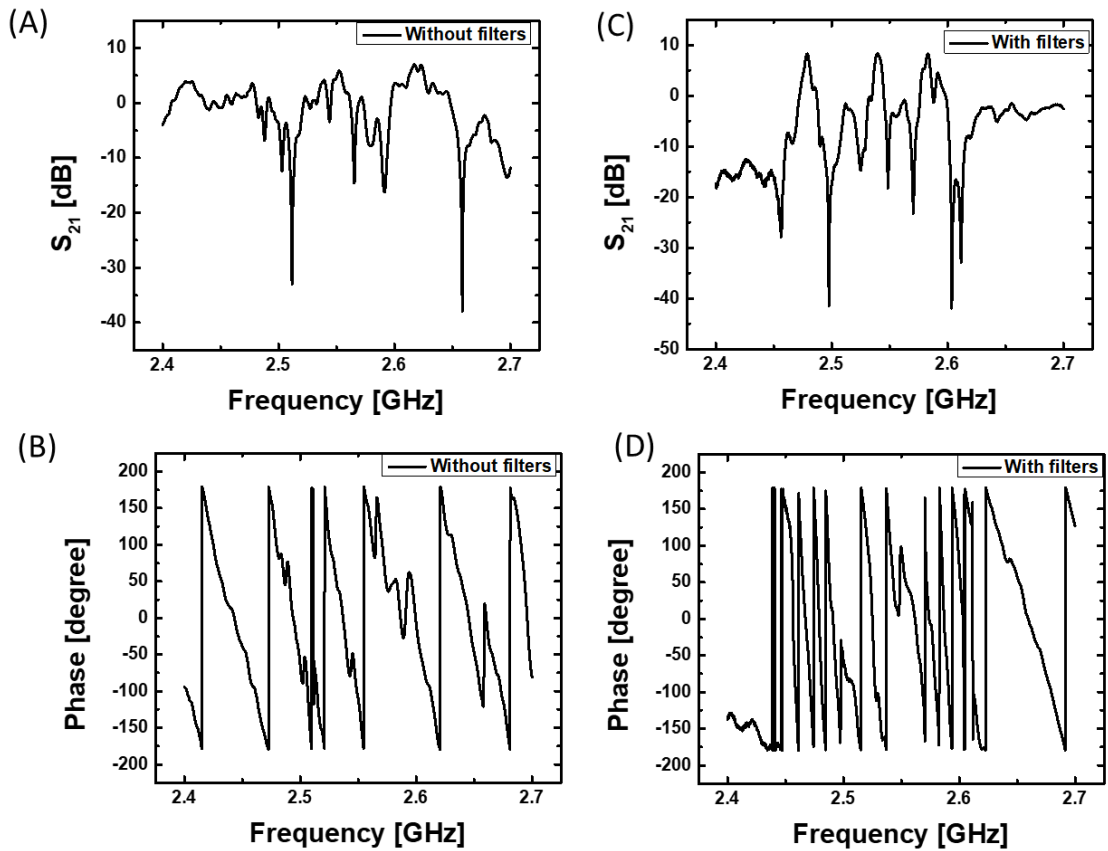

Figure [s4]: Experimental data obtained with PNA showing spin wave transmission through the film. The bias magnetic field is about 375 Oe and directed in-plane on the film surface. Plots (A) and (B) show S21 and phase shift as a function of frequency for the film without frequency filters. Plots (C) and (D) show S21 and phase shift as a function of frequency for the film frequency filters.

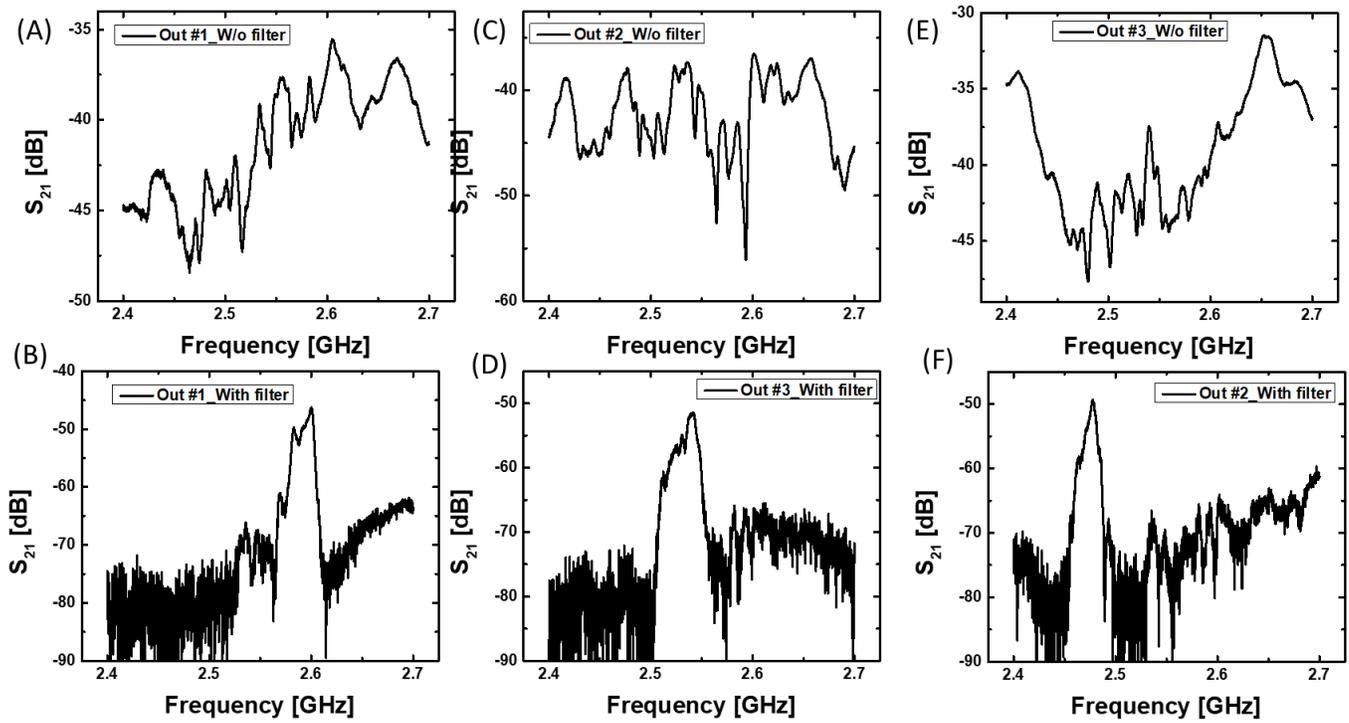

Figure [s5]: Experimental data obtained with PNA showing spin wave transmission through the film. The bias magnetic field is about 375 Oe and directed in-plane on the film surface. Plots (A) and (B) show S21 and phase shift as a function of frequency for output #1 without and with the frequency filter. Plots (C) and (D) show S21 and phase shift as a function of frequency for output #2 without and with the frequency filter. Plots (E) and (F) show S21 and phase shift as a function of frequency for output #2 without and with the frequency filter.

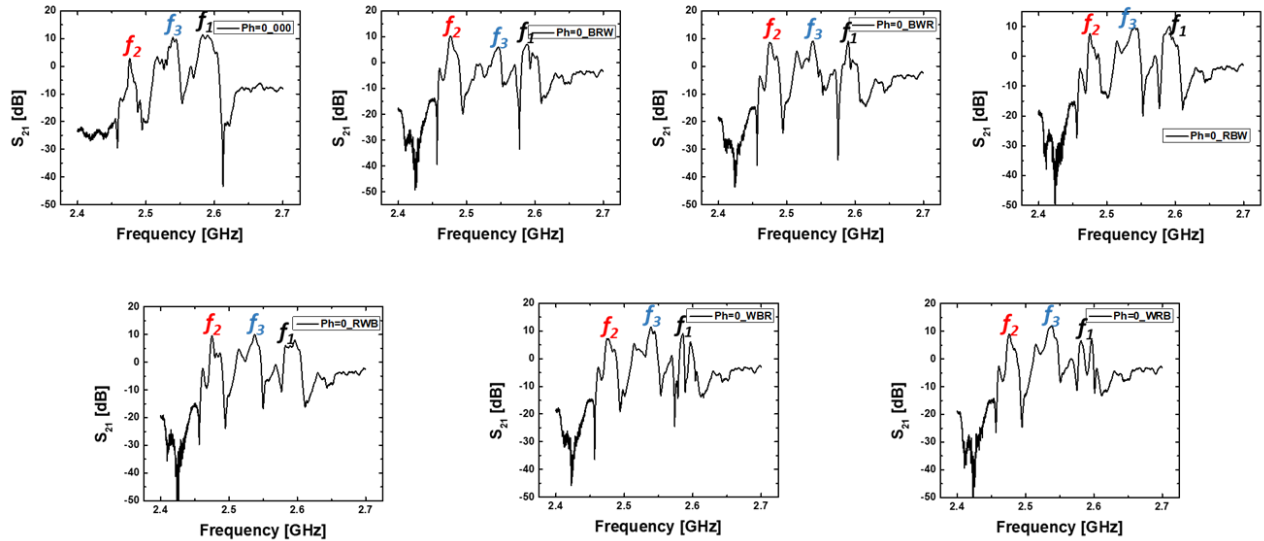

Figure [s6]: Experimental data: Auto-Oscillations in the Active Ring Circuit. Magnetic mesh is connected to the electric part. Auto-oscillations are observed for certain levels of amplification and the external phase shifter. The gain in the circuit is set to +13.5 dB.
(a) Frequencies of the auto-oscillations depending on the external phase shifter. One div is equivalent to $\pi/30$ radians (b) Total power in the active ring circuit depending on the external phase. (c) Power of the auto-oscillations in different propagation routes.